\documentclass[a4paper,12pt]{article}
\usepackage{graphicx}
\usepackage[footnotesize]{caption}

\usepackage{multirow}
\usepackage{dsfont}
\usepackage{amssymb}
\usepackage{slashed}
\usepackage{mathtools}
\usepackage{nicefrac}
\usepackage{verbatim}

\usepackage{axodraw}
\usepackage{color}

\newcommand{\rr}[1]{\mathrm{#1}}
\newcommand{\cc}[1]{\mathcal{#1}}

\newcommand{\dd}[1]{\mathds{#1}}
\newcommand{\ii}[1]{\widetilde{#1}}

\newcommand{\essm}{E$_6$SSM }

\def\beq{\begin{equation}}
\def\eeq{\end{equation}}
\newcommand{\be}{\begin{eqnarray}}
\newcommand{\nn}{\nonumber}
\newcommand{\ee}{\end{eqnarray}}
\newcommand{\ba}{\begin{array}}
\newcommand{\ea}{\end{array}}

\newcommand{\f}[2]{\frac{#1}{#2}}
\newcommand{\nf}[2]{\nicefrac{#1}{#2}}

\begin{document}

\begin{titlepage}
\phantom{~}
\vskip30mm
\begin{center}
{\sffamily\Large 
NMSSM+
}
\\[8mm]
Jonathan~P.~Hall $^{1,}$\footnote{halljp@indiana.edu}
and Stephen~F.~King $^{2,}$\footnote{king@soton.ac.uk}
\\[3mm]
{\small\it $^{1}$ Physics Department, Indiana University,\\
                  Bloomington, IN 47405, USA}\\[1mm]
{\small\it $^{2}$ School of Physics \& Astronomy, University of Southampton,\\
                  Highfield, Southampton SO17 1BJ, UK}\\[1mm]
\end{center}
\vspace*{20mm}

\begin{abstract}
\noindent
It is well known that the scale invariant NMSSM has lower fine-tuning than the MSSM, but suffers from
the domain wall problem.
We propose a new improved scale invariant version of the NMSSM, called the NMSSM+,
which introduces extra matter
in order to reduce even more the fine-tuning of the NMSSM.
The NMSSM+ also provides 
a resolution of the domain wall problem of the NMSSM due to a discrete $R$-symmetry, 
which also stabilises the proton.
The extra matter descends from an E$_6$ gauge group
and fills out three complete 27-dimensional representations at the TeV scale,
as in the E$_6$SSM. However the $U(1)_N$ gauge group of the E$_6$SSM is broken at a high energy scale 
leading to reduced fine-tuning.
The extra matter of the NMSSM+ includes charge $\pm1/3$ colour triplet D-fermions 
which may be naturally heavier than the weak scale because they 
receive their mass from singlet field vacuum expectation values
other than the one responsible for the weak scale effective $\mu$ parameter.
\end{abstract}

\end{titlepage}
\newpage
\setcounter{footnote}{0}

\section{Introduction}
The recent discovery of a candidate Higgs-like boson
with a mass around
$\sim125$--126~GeV~\cite{:2012gk,:2012gu}
is tremendously exciting since it may provide
a window into new physics Beyond the Standard Model (BSM),
for example Supersymmetry
(SUSY)~\cite{Higgs,Hall:2011aa,Arvanitaki:2011ck,Ellwanger:2011aa,Gunion:2012zd}.

In the Minimal Supersymmetric Standard Model (MSSM) the lightest Higgs
boson is lighter than about 130--135~GeV, depending on top squark
parameters (see e.g.~\cite{Djouadi:2005gj} and references therein).
A 125~GeV SM-like Higgs boson is consistent with the MSSM in the
decoupling limit. In the limit of decoupling the light Higgs mass is given by
\begin{equation}
m_h^2 \, \approx  \, M_Z^2 \cos^2 2 \beta + \Delta m_h^2, 
\label{eq:hmassMSSM}
\end{equation}
where $ \Delta m_h^2$ is dominated by loops of heavy top quarks and
top squarks and $\tan \beta$ is the ratio of the vacuum expectation
values (VEVs) of the two Higgs doublets introduced in the MSSM Higgs
sector. At large $\tan \beta$ we require $\Delta m_h \approx 85$~GeV,
which means that a very substantial loop contribution, nearly as large
as the tree-level mass, is needed to raise the Higgs boson mass to 125~GeV,
leading to some degree of fine-tuning.

In the light of such fine-tuning considerations, it has been known for some time, even after the 
LEP limit on the Higgs boson mass of 114 GeV, that the fine-tuning of the
MSSM could be ameliorated in the scale invariant Next-to-Minimal Supersymmetric
Standard Model (NMSSM)~\cite{NMSSMtuning}. With a 125 GeV Higgs boson,
this conclusion is greatly strengthened and the NMSSM appears to be a
much more natural alternative. In the NMSSM, the spectrum of the MSSM is extended by
one singlet superfield~\cite{genNMSSM1,genNMSSM2,other-non-minimal,Nevzorov:2004ge}
(for reviews see~\cite{Ellwanger:2009dp,Maniatis:2009re}). In the NMSSM the
supersymmetric Higgs mass parameter $\mu$ is promoted to a
gauge-singlet superfield, $S$, with a coupling to the Higgs doublets,
$\lambda S H_d H_u$, that is perturbative up to unified scales.
The maximum mass of the lightest Higgs boson is  
\begin{equation}
m_h^2 \, \approx  \,  M_Z^2 \cos^2 2 \beta + \f{\lambda^2 v^2}{2} \sin^2 2 \beta +  \Delta m_h^2,
\label{eq:hmassNMSSM}
\end{equation}
where here $v=246$~GeV.  For $\lambda v > M_Z$, the tree-level
contributions to $m_h$ are maximized for moderate values of $\tan \beta$ 
rather than by large values of $\tan \beta$ as in the MSSM. 
For example, taking
$\lambda =0.6$ and $\tan\beta=2$, these tree-level
contributions raise the Higgs boson mass to about 100~GeV, 
and $\Delta m_h \sim 75$~GeV is required 
to achieve a Higgs mass of 125~GeV. This is to be compared 
to the MSSM requirement $\Delta m_h \gtrsim 85$~GeV. The difference 
between these two values (numerically about 10~GeV) is significant since
$\Delta m_h$ depends logarithmically on the stop
masses as well as receiving an important contribution from stop
mixing.

In the NMSSM 
$\lambda \sim 0.6$ is the largest value in order not to spoil the validity of perturbation 
theory up to the GUT scale for $\tan\beta\sim 2$ 
and for $k\sim 0.2$ (where $\lambda $ and $k$ are evaluated at $M_Z$)~\cite{King:2012is}. 
However it has been known for some time~\cite{Masip:1998jc} that the presence of additional extra matter allows 
larger values of $\lambda $ to be achieved.
For example, adding three families of $5+\overline 5$ extra matter at a mass scale of 300~GeV 
increases the largest value to $\lambda \sim 0.7$, for the same parameters as before~\cite{King:2012is}.
For example, taking
$\lambda \sim 0.7$ and $\tan\beta \sim 2$, these tree-level
contributions raise the Higgs boson mass to about 112~GeV, 
and $\Delta m_h \sim 55$~GeV is required 
to achieve a Higgs mass of 125 GeV. This is to be compared to 
$\Delta m_h \sim 75$~GeV required for $\lambda =0.6$.
The difference between these two values (numerically about 20~GeV) is more
than the difference between the MSSM and the NMSSM, and 
can lead to a further significant reduction in fine-tuning.
The above discussion shows that there is an argument from fine-tuning
for extending the NMSSM to include extra matter. 

An example of a model with extra matter is the Exceptional Supersymmetric Standard Model 
(E$_6$SSM)~\cite{King:2005jy,King:2005my,King:2007uj,Athron:2009bs,Athron:2009ue,Athron:2011wu,Accomando:2006ga,Hall:2010ix,Belyaev:2012si},
where 
the spectrum of the MSSM is extended to fill out three complete 27-dimensional representations
of the gauge group E$_6$ which is broken at the unification scale down to the SM gauge group
plus an additional gauged $U(1)_N$ symmetry at low energies under which right-handed neutrinos are neutral,
allowing them to get large masses.
The three $27_i$-plet families 
(labelled by $i=1,2,3$) contain the usual quarks and leptons plus the following extra states:
SM-singlet fields, $S_i$; up- and down-type Higgs doublets, $H_{ui}$ and $H_{di}$;
and charged $\pm 1/3$ coloured, exotics $D_i$ and $\bar{D}_i$.
The extra matter ensures anomaly cancellation, however
the model also contains two extra SU(2) doublets, $H'$ and $\bar{H}'$, which are
required for gauge coupling unification~\cite{King:2007uj}.
To evade rapid proton decay a $\dd{Z}_2$ symmetry, either $\dd{Z}_2^{qq}$ or $\dd{Z}_2^{lq}$,
is introduced and to evade large flavour changing neutral currents an
approximate $\dd{Z}_2^H$ symmetry is introduced which ensures that
only the third family of Higgs doublets $H_{u3}$ and $H_{d3}$ couple to
fermions and get vacuum expectation values (VEVs).
Similarly only the third family singlet $S_3$ gets a VEV, $\langle S_3 \rangle = s/\sqrt{2}$,
which is responsible for the effective $\mu$ term and D-fermion masses.
The first and second families of Higgs doublets and SM-singlets,
which do not get VEVs, are called ``inert''.
The maximum mass of the lightest SM-like Higgs boson in the E$_6$SSM is~\cite{King:2005jy}
\begin{equation}
m_h^2 \, \approx  \,  M_Z^2 \cos^2 2 \beta + \f{\lambda^2 v^2}{2} \sin^2 2 \beta +
\frac{M_Z^2}{4} \left(1+ \frac{1}{4}\cos 2 \beta  \right)^2 + \Delta m_h^2,
\label{eq:hmassE6SSM}
\end{equation}
where the extra contribution relative to the NMSSM value
in equation~(\ref{eq:hmassNMSSM}) is due to the $U(1)_N$ $D$-term.
The Higgs mass can be larger due to two separate reasons; firstly the value of $\lambda$ may be larger
due to the extra matter and secondly there is a $U(1)_N$ $D$-term contribution equal to
$\frac{1}{2}M_Z$ ($\frac{3}{8}M_Z$)~GeV
for low (high) $\tan \beta$. For example for $\tan \beta \sim 2$, and $\lambda \sim 0.7$,
the E$_6$SSM can completely account for a Higgs boson mass of 125 GeV at the tree-level,
without the need for any extra contribution from radiative corrections, i.e. with $\Delta m_h \sim 0$.

Although the $D$-term from the low energy $U(1)_N$ gauge group
appears to help with fine-tuning, by increasing the tree-level Higgs mass,
in fact it leads to a new fine-tuning problem associated with the non-observation
of the $Z'_N$ gauge boson. The reason is that the singlet VEV $s$, which is responsible for both the
effective $\mu$ term and the $Z'_N$ mass, must be quite large, in practice $s>5$~TeV for 
$M_{Z'_N}>2$~TeV, which is the current experimental limit~\cite{Chatrchyan:2012it}. 
As we shall see, such a large singlet VEV is unnatural if that singlet is
responsible for the effective $\mu$ term,
due to singlet $D$-terms entering the Higgs potential. Furthermore, a large singlet VEV implies
a large effective $\mu$ term, at least for non-small $\lambda$, which also leads to fine-tuning.
It might appear that if $\lambda$ is small then the effective $\mu$ term may be made small so that
we are back to the fine-tuning situation of the MSSM.
However, as we shall see, the tree-level potential involves the
$Z'_N$ mass squared explicitly and can only be balanced by the effective $\mu$-squared term,
leading to large and unavoidable tree-level fine-tuning.

In this paper we propose a model called the NMSSM+, 
containing the extra matter content of the E$_6$SSM,
but without a low energy $U(1)_N$ gauge group, this being broken close to the unification scale
by a high energy mechanism which 
does not give rise to mass for any of the components of the $27_i$ states above.
The absence of the $U(1)_N$ gauge group 
immediately removes all related $D$-terms from the low energy theory, and we return to a
situation similar to the NMSSM with extra matter where fine-tuning may be lowered
as discussed above, without encountering any new fine-tuning problems due to the
large singlet VEV. In order to achieve the smallest fine-tuning possible,
we want to lower the singlet VEV $s$ as much as possible, while maintaining
a large value of $\lambda$ in order to increase
the tree-level Higgs mass, so that the effective $\mu$ term is as small as possible.
If only 
the third generation SM-singlet acquires
a VEV, $\langle S_3\rangle = s_3/\sqrt{2} = s/\sqrt{2}$,
then this implies that the coloured exotics $D_i$, $\bar{D}_i$
should also have low masses of order $\mu$, and their non-observation may imply a lower
limit on $\mu$, leading again to increased fine-tuning.
In order to avoid this, unlike in the E$_6$SSM
we shall allow the first and second family SM-singlets also to 
acquire VEVs $\langle S_{\alpha}\rangle = s_{\alpha}/\sqrt{2}$.
To minimise the fine-tuning associated with the $\mu$ parameter we 
assume $s_{\alpha}\gg s$ with $s$ being responsible for the smaller effective $\mu$ term,
and $s_{\alpha}$ being responsible for the larger masses of the $D_i$, $\bar{D}_i$ exotics.
Another difference from the E$_6$SSM is that we shall generate cubic singlet interactions
$S_i^3$ as in the NMSSM (for all three singlets $i=1,2,3$), which breaks the Peccei-Quinn (PQ) symmetry. 
The associated domain
wall problem which arises when the accompanying discrete $\dd{Z}_3$ symmetry is broken will
be avoided in our model, however.
This is due to the fact that in the NMSSM+ instead of $\dd{Z}_2^{qq}$ or $\dd{Z}_2^{lq}$ we impose either
$\dd{Z}_4^{qq}$ or $\dd{Z}_4^{lq}$. These are $R$-symmetries under which the superpotential itself transforms.
Note that, although all three families of singlets acquire VEVs, 
only the third generation of Higgs doublets
acquire VEVs $\langle H^0_{d3}\rangle = v_d/\sqrt{2}$ and
$\langle H^0_{u3}\rangle = v_u/\sqrt{2}$,
as in the E$_6$SSM.

The layout of the remainder of the paper is as follows: In Section~2 we briefly review the NMSSM and E$_6$SSM
and give an overview of the NMSSM+. This treatment may be sufficient for those readers only interested in the 
phenomenology of the NMSSM+.
In Section~3 we summarise
the superpotential and symmetries of the high energy NMSSM+, which is similar to the 
E$_6$SSM but contains a $U(1)_N$ breaking sector and somewhat different symmetries,
and show how it leads to the low energy  NMSSM+ outlined in Section~2. In Section~4 
we discuss the important aspects of the NMSSM+ and explain the 
reason for and implications of the various symmetries in detail.
The tree-level fine-tuning problem of the E$_6$SSM is explained in Section~5
and the lower expected fine-tuning in the NMSSM+ is explained. The concluding section is
Section~6.

\section{Comparing the NMSSM, E$_6$SSM, and NMSSM+}

\subsection{Overview of the NMSSM}

The renormalisable, scale invariant superpotential of the NMSSM~\cite{genNMSSM1,genNMSSM2} is
\be
W^{\rr{NMSSM}} &=& \lambda\hat{S}\hat{H}_{d}\hat{H}_{u} + 
 \f{k}{3}\hat{S}^3 + W_{\rr{Yukawa}},
 \label{NMSSM}
\ee
where $W_{\rr{Yukawa}}$ is the usual MSSM-like Yukawa superpotential terms involving Higgs doublets
\be
W_{\rr{Yukawa}} &=& h^N_{ij}\hat{H}_{u}\hat{L}_{Li}\hat{N}^c_j + h^U_{ij}\hat{H}_{u}\hat{Q}_{Li}\hat{u}_{Rj}^c
+ h^D_{ij}\hat{H}_{d}\hat{Q}_{Li}\hat{d}_{Rj}^c + h^E_{ij}\hat{H}_{d}\hat{L}_{Li}\hat{e}_{Rj}^c.
\label{yuk}
\ee
The gauge symmetry is that of the SM and the superfield $\hat{S}$ is 
a complete gauge singlet. When the scalar component of $\hat{S}$ acquires a
VEV this VEV generates an effective $\mu$ term, coupling the Higgs
doublets $\hat{H}_{(d,u)}$.
Alternative non-scale invariant models known as the Minimal
Non-minimal Supersymmetric SM (MNSSM), the new minimally-extended supersymmetric SM,
the nearly-Minimal Supersymmetric SM (nMSSM), and the generalised 
NMSSM have been considered elsewhere \cite{other-non-minimal}.
By contrast in this paper the extension of the NMSSM we consider,
namely the NMSSM+,  will be scale invariant to very good approximation.

\subsection{Overview of the \essm}

We first recall that the
E$_6$SSM~\cite{King:2005jy,King:2005my,King:2007uj,Athron:2009bs,Athron:2009ue,Athron:2011wu,Accomando:2006ga,Hall:2010ix,Belyaev:2012si} 
may be derived from an 
$E_6$ GUT group broken via the following symmetry breaking chain:
\be
E_6 &\rightarrow& SO(10) \otimes U(1)_\psi \nn\\
&\rightarrow& SU(5) \otimes U(1)_\chi \otimes U(1)_\psi \nn\\
&\rightarrow& SU(3) \otimes SU(2) \otimes U(1)_Y
\times U(1)_\chi \otimes U(1)_\psi \nn\\
&\rightarrow& SU(3) \otimes SU(2) \otimes U(1)_Y \otimes U(1)_N.
\ee
In practice it is assumed that the above symmetry breaking chain 
occurs at a single GUT scale $M_X$ in one step,
due to some unspecified symmetry breaking sector,
\be
E_6 &\rightarrow& SU(3) \otimes SU(2) \otimes U(1)_Y \otimes U(1)_N, 
\ee
where
\be
U(1)_N &=& \cos(\vartheta) U(1)_\chi + \sin(\vartheta) U(1)_\psi
\ee
and $\tan(\vartheta) = \sqrt{15}$ such that the right-handed neutrinos
that appear in the model are completely
chargeless. The $U(1)_N$ gauge group remains unbroken down to the TeV energy scale.
Three complete 27 representations of $E_6$ then also must survive
down to the TeV scale in order to ensure anomaly cancellation. These $27_i$ 
decompose under the $SU(5) \otimes U(1)_N$ subgroup as follows:
\be
27_i &\rightarrow& (10,1)_i + (\bar{5},2)_i + (\bar{5},-3)_i + (5,-2)_i
+ (1,5)_i + (1,0)_i,
\ee
where the $U(1)_N$ charges must be GUT normalised by a factor of $1/\sqrt{40}$.
The first two terms contain the usual quarks and leptons, whereas the final term, which is
a singlet under the entire low energy gauge group, contains the
(CP conjugated) right-handed
neutrinos ${N}^c_i$. The second-to-last term, which is charged only under $U(1)_N$,
contains
the SM-singlet fields $S_i$. 
The remaining terms $(\bar{5},-3)_i $ and $(5,-2)_i$
contain three families of 
up- and down-type Higgs doublets, $H_{ui}$ and $H_{di}$,
and charged $\pm 1/3$ coloured exotics, $D_i$ and $\bar{D}_i$.
These are all superfields written with hats below.

The low energy gauge invariant superpotential can be written
\be
W^{\rr{E}_6\rr{SSM}} &=& W_0 + W_{1,2},\label{eq:w}
\ee
where $W_{0,1,2}$ are given by
\be
W_0 &=& \lambda_{ijk}\hat{S}_i\hat{H}_{dj}\hat{H}_{uk} + \kappa_{ijk}\hat{S}_i\hat{\bar{D}}_j\hat{D}_k + h^N_{ijk}\hat{N}^c_i\hat{H}_{uj}\hat{L}_{Lk}\label{eq:w0}\nn\\
&&\phantom{~} + h^U_{ijk}\hat{H}_{ui}\hat{Q}_{Lj}\hat{u}_{Rk}^c + h^D_{ijk}\hat{H}_{di}\hat{Q}_{Lj}\hat{d}_{Rk}^c + h^E_{ijk}\hat{H}_{di}\hat{L}_{Lj}\hat{e}_{Rk}^c,
\label{W0}\\
W_1 &=& g^Q_{ijk}\hat{D}_i\hat{Q}_{Lj}\hat{Q}_{Lk} + g^q_{ijk}\hat{\bar{D}}_i\hat{d}_{Rj}^c\hat{u}_{Rk}^c,\label{eq:w1}\\
W_2 &=& g^N_{ijk}\hat{N}^c_i\hat{D}_j\hat{d}_{Rk}^c + g^E_{ijk}\hat{D}_i\hat{u}_{Rj}^c\hat{e}_{Rk}^c +
 g^D_{ijk}\hat{\bar{D}}_i\hat{Q}_{Lj}\hat{L}_{Lk},\label{eq:w2}
\ee
with $W_{1,2}$ referring to either $W_1$ or $W_2$.

If one neglects the $E_6$ violating bilinear terms $\hat{H}_{ui}\hat{L}_{Lj}$ and $\hat{D}_i\hat{d}_{Rj}^c$
then one can see that at the renormalisable level the gauge invariance ensures
matter parity and hence LSP stability. All lepton and quark superfields are defined to be odd under
matter parity $\dd{Z}_2^M$, while $\hat{H}_{ui}$, $\hat{H}_{di}$,  $\hat{D}_i$,
$\hat{\bar{D}}_i$, and $\hat{S}_i$ are even.
This means that the fermions associated with $\hat{D}_i$, $\hat{\bar{D}}_i$ are
SUSY particles analogous to the Higgsinos, while their scalar components may be thought of as colour-triplet
(and electroweak singlet) Higgses, making complete $5$ and $\bar{5}$ representations
without the usual doublet-triplet splitting.

In order for baryon and
lepton number to also be conserved, preventing rapid proton decay mediated by
$\hat{D}_i$, $\hat{\bar{D}}_i$, one imposes either
$\dd{Z}_2^{qq}$ or $\dd{Z}_2^{lq}$.
Under the former the lepton, including the RH neutrino, superfields are odd
and under the latter both
the lepton and the $\hat{D}_i$, $\hat{\bar{D}}_i$ superfields are odd. Under the former $W_2$ is forbidden and
under the latter $W_1$ is forbidden. Baryon and lepton number are then both conserved
at the renormalisable level,
with the $\hat{D}_i$, $\hat{\bar{D}}_i$ interpreted as diquarks in the former case and leptoquarks
in the latter case.

A further approximate flavour symmetry $\dd{Z}_2^H$ is also assumed. It is this approximate
symmetry that distinguishes the third (by definition, ``active'') generation of Higgs doublets and
SM-singlets from the second and first (``inert'') generations. Under this approximate symmetry
all superfields other than the active
$\hat{S}=\hat{S}_3$, $\hat{H}_d=\hat{H}_{d3}$, and $\hat{H}_u=\hat{H}_{u3}$ are odd. The
inert fields then have small couplings to matter and do not radiatively acquire VEVs or
lead to large flavour changing neutral currents. The active fields can have large couplings to
matter and radiative electroweak symmetry breaking (EWSB) occurs with these fields.
In particular the multi-TeV scale VEV $\langle S \rangle = \langle S_3 \rangle = s/\sqrt{2}$
is responsible for breaking the $U(1)_N$ gauge group and generating the
effective $\mu$ term and D-fermion masses. In particular we must have $s>5$~TeV in order to satisfy 
$M_{Z'_N}>2$~TeV, which is the current experimental limit~\cite{Chatrchyan:2012it},
leading to large fine-tuning.

\subsection{Overview of the NMSSM+}

At high energies, just below the GUT scale, the NMSSM+ theory includes the matter content and gauge group
of the E$_6$SSM, including the $U(1)_N$ gauge group.
In order to break the $U(1)_N$ gauge group we introduce a renormalisable
term $\hat{\Sigma}(l\hat{\cc{S}}\hat{\bar{\cc{S}}} - M_\Sigma^2)$, 
where the two extra SM-singlet superfields
$\hat{\cc{S}}$ and $\hat{\bar{\cc{S}}}$ have $U(1)_N$ charges $Q^N_S$ and $-Q^N_S$
respectively (where $Q^N_S$ is also the $U(1)_N$ charge of the usual E$_6$SSM SM-singlets $S_i$)
and $\hat{\Sigma}$ is 
a complete $G_{\rr{SM}} \otimes U(1)_N$ singlet superfield 
with $l$ being a dimensionless Yukawa coupling constant.
This superpotential breaks the $U(1)_N$ at the
intermediate scale $M_\Sigma$~\footnote{It has not escaped our attention
that this $U(1)_N$ symmetry breaking sector may provide
an example of
SUSY hybrid inflation with the scalar component of $\hat{\Sigma}$ being the inflaton and those of
$\hat{\cc{S}}$ and $\hat{\bar{\cc{S}}}$ being the waterfall fields~\cite{Dvali:1994ms,BasteroGil:2006cm}.}.
Since $\hat{\cc{S}}$ has the same gauge charges as $\hat{S_i}$,
we propose that the $R$-symmetry of the theory is used to forbid superpotential
terms such as $\hat{\cc{S}}\hat{H}_d.\hat{H}_u$
and $\hat{\Sigma} \hat{S}\hat{\bar{\cc{S}}}$
(where $\hat{\cc{S}}$ takes the place of $\hat{S}_i$ or vice versa) so that
the hierarchy between the $\Sigma$ scale and the EWSB scale can be naturally maintained.
In addition, various non-renormalisable terms are also included, all controlled by extra symmetries
as discussed in the next section. In particular, some of the non-renormalisable terms yield
low energy cubic singlet couplings of the form $\hat{S}_i\hat{S}_j\hat{S}_k$, including the 
NMSSM cubic singlet coupling.

At low energies, the scale invariant NMSSM+ contains the matter and Higgs content of three
27 dimensional superfield representations of $E_6$,
minus the three RH neutrinos $\hat{N}_{i}$ which being complete singlets may get very large masses,
leaving the three quark and lepton families,
$\hat{Q}_{Li}$, $\hat{u}_{Ri}$, $\hat{d}_{Ri}$, $\hat{L}_{Li}$,
$\hat{e}_{Ri}$;
three families of Higgs doublets, $\hat{H}_{di}$ and
$\hat{H}_{ui}$; three families of colour triplet and antitriplet states,
$\hat{D}_{i}$ and $\hat{\bar{D}}_{i}$;
and three SM-singlets, $\hat{S}_{i}$, where we define $ \hat{S}= \hat{S}_{3}$
and $ \hat{S}_{\alpha}= \hat{S}_{1,2}$, and similarly for the Higgs doublets.
The low energy superpotential of the NMSSM+ obeying all of the symmetries of the model
is approximately that of the NMSSM {\em plus} an extra sector,
\be
\label{NMSSM+}
W^{\rr{NMSSM+}} &\approx& W^{\rr{NMSSM}} + W^{\rr{extra}},
\ee
where $W^{\rr{NMSSM}}$ is the same as the NMSSM superpotential in equation~(\ref{NMSSM}),
while $ W^{\rr{extra}}$ includes the extra terms associated with the couplings of the extra two families of 
Higgs doublets and singlets and three families of colour triplets, 
\be
\label{exotic}
W^{\rr{extra}} &\approx&  \lambda_{\alpha\beta\gamma}\hat{S}_\alpha\hat{H}_{d\beta}\hat{H}_{u\gamma}
+ \kappa_{\alpha ij}\hat{S}_\alpha\hat{\bar{D}}_i\hat{D}_j
+ \f{k_{\alpha\beta\gamma}}{3}\hat{S}_\alpha\hat{S}_\beta\hat{S}_\gamma + W_{1,2},
\label{Wextra}
\ee
where $i,j,k\in\{1,2,3\}$ whereas $\alpha,\beta,\gamma\in\{1,2\}$.
We have neglected couplings (other than those in $W_{1,2}$)
that are suppressed under an approximate $\dd{Z}_3^{HD}$ symmetry.
$\hat{S}=\hat{S}_3$ is responsible for the effective
Higgs $\mu$ term of the active third generation Higgs
doublets $\hat{H}_{(d,u)}=\hat{H}_{(d,u)3}$ ($\mu=\lambda \langle S\rangle$),
while $\hat{S}_{1,2}$
is responsible for similar effective $\mu$ terms for the inert generations
of Higgs doublets ($\mu_{\beta\gamma}
= \lambda_{\alpha\beta\gamma}\langle S_\alpha\rangle$) and for the induced D-fermion masses.
Note that we expect all three SM-singlets to develop VEVs,
with $\langle S_3\rangle \ll \langle S_\alpha \rangle$ in
order to allow a relatively small $\mu$ term (and low fine-tuning)
and relatively large exotic $\hat{D}$ and $\hat{\bar{D}}$ particle masses.

As mentioned in the Introduction,
in the NMSSM+ instead of $\dd{Z}_2^{qq}$ or $\dd{Z}_2^{lq}$ we impose either
$\dd{Z}_4^{qq}$ or $\dd{Z}_4^{lq}$. These are $R$-symmetries under which the superpotential itself transforms.
All terms in whichever of $W_{1,2}$ is allowed
by the $\dd{Z}_4$ $R$-symmetry will be suppressed under an approximate $\dd{Z}_3^{HD}$
symmetry, as discussed in the next section and summarised in Table~\ref{sym}.
Nonetheless, the suppressed terms in whichever of $W_{1,2}$
is allowed by the $R$-symmetry
will allow the exotic, coloured $\hat{D}$ and $\hat{\bar{D}}$ particles to decay.

\section{The NMSSM+ }
\label{model}

We first define the high energy symmetry and superpotential of the NMSSM+,
then give the resulting scale invariant low energy effective theory relevant for phenomenology.

\subsection{The high energy NMSSM+ }
As discussed above, the high energy NMSSM+ 
includes the superfield content and gauge group of the \essm (including the $U(1)_N$ gauge group) plus the renormalisable 
term $\hat{\Sigma}(l\hat{\cc{S}}\hat{\bar{\cc{S}}} - M_\Sigma^2)$, which spontaneously breaks $U(1)_N$
at the scale $M_\Sigma < M_X$,
plus some other non-renormalisable terms,
all controlled by a set of symmetries. In this section we give a precise definition of the model,
including the non-renormalisable terms and the full set of symmetries.

The full model, valid at a high energy scale just below the GUT breaking scale $M_X$, is defined by the 
superfields and symmetries given in Table~\ref{sym}.
The resulting gauge invariant high energy superpotential, including important non-renormalisable terms, is given by,
\be
\cc{W}^{\rr{NMSSM+}} &=& W^{\rr{E}_6\rr{SSM}}+ W^{\cc{S}}+ \Delta W_0 + \Delta W_1,
\label{NMSSM+3}
\ee
where $W^{\rr{E}_6\rr{SSM}}$ is the superpotential of the \essm given in equation~(\ref{eq:w}), while
the remaining parts involve the two extra SM-singlet superfields
$\hat{\cc{S}}$ and $\hat{\bar{\cc{S}}}$ with $U(1)_N$ charges $Q^N_S$ and $-Q^N_S$
respectively, where $Q^N_S$ is also the $U(1)_N$ charge of the usual E$_6$SSM SM-singlets $S_i$,
\be
\label{wS}
W^{\cc{S}} &=&\hat{\Sigma}\left(l\hat{\cc{S}}\hat{\bar{\cc{S}}} - M_\Sigma^2\right)\nn\\
&&\phantom{~} + \f{b_{ijk}}{M_X^3}\hat{S}_i\hat{S}_j\hat{S}_k\hat{\bar{\cc{S}}}^3
+ \f{d_{ijk}}{M_X}\hat{S}_i\hat{N}^c_j\hat{N}^c_k\hat{\bar{\cc{S}}}\nn\\
&&\phantom{~} + \left[\f{c}{M_X^{11}}\hat{S}^7\hat{\bar{\cc{S}}}^7
+ \f{c'}{M_X^8}\hat{S}^5\hat{H}_d.\hat{H}_u\hat{\bar{\cc{S}}}^4
+ \f{c''}{M_X^5}\hat{S}^3(\hat{H}_d.\hat{H}_u)^2\hat{\bar{\cc{S}}}+\cdots\right],\\
\label{delw0}
\Delta W_0 &=& \f{r^{udd}_{ijk}}{M_X}\hat{u}_{Ri}^c\hat{d}_{Rj}^c\hat{d}_{Rk}^c\hat{\bar{\cc{S}}} + \f{r^{DDu}_{ijk}}{M_X}\hat{\bar{D}}_i\hat{\bar{D}}_j\hat{u}_{Rk}^c\hat{\cc{S}},\\
\label{delw1}
\Delta W_1 &=& \f{r^{SDd}_{ijk}}{M_X}\hat{S}_i\hat{D}_j\hat{d}_{Rk}^c\hat{\bar{\cc{S}}} + \f{r^{HDq}_{ijk}}{M_X}\hat{H}_{di}\hat{\bar{D}}_j\hat{Q}_{Lk}\hat{\bar{\cc{S}}}.
\ee

The r\^ole of the non-renormalisable terms in $W^{\cc{S}}$ is as follows.
The VEVs for $\cc{S}$ and $\bar{\cc{S}}$ at the scale $M_\Sigma$, as well as 
breaking $U(1)_N$, induce desirable cubic SM-singlet terms (required for the NMSSM)
through the non-renormalisable couplings proportional to $b_{ijk}$.
The $\cc{S}$ and $\bar{\cc{S}}$ VEVs also induce SM-singlet couplings to RH neutrinos at low energy
controlled by $d_{ijk}$, however these terms may give negligible contributions to RH neutrino mass
compared to other sources of mass. 
The non-renormalisable terms with 7 conventional low energy fields in the square brackets are
important for solving domain wall problems,
as explained in subsection~\ref{domain}.

The matter parity $\dd{Z}_2^M$ actually forbids
the non-renormalisable terms in both $\Delta W_0$ and $\Delta W_1$,
but we have included them in order to see the effect of relaxing $\dd{Z}_2^M$,
as considered in subsection~\ref{RPV}.
Both these terms yield $R$-parity violating terms not present in the \essm
(since $R$-parity is automatically conserved by the renormalisable E$_6$SSM superpotential) 
once the VEVs for $\cc{S}$ and $\bar{\cc{S}}$ are inserted.
Without $\dd{Z}_4^{qq}$ or $\dd{Z}_4^{lq}$ other R-parity violating terms would also appear.
Depending on which of the two options, namely $\dd{Z}_4^{qq}$ or $\dd{Z}_4^{lq}$, is chosen
either $W_2$ is forbidden or both $W_1$ and $\Delta W_1$ are forbidden, respectively,
forbidding rapid proton decay
(see subsection~\ref{RPV}).

\begin{table}
\begin{center}
\begin{tabular}{c||c|c|c|c||cc|c|c|}
& $SU(3)$ & $SU(2)$ & $U(1)_Y$ & $U(1)_N$ & \multicolumn{2}{c|}{either} & optional & approx. \\
& rep. & rep. & $\sqrt{5/3}Q^{Y}$ & $\sqrt{40}Q^{N}$ &
$\dd{Z}_4^{qq}$ & $\dd{Z}_4^{lq}$ & $\dd{Z}_2^M$ & $\dd{Z}_3^{HD}$ \\\hline
$\hat{Q}_{Li}$ & $3$ & $2$ & $+\nf{1}{6}$ & $+1$ & $+i$ & $+i$ & $-1$ & +1 \\
$\hat{d}^{c}_{Ri}$ & $\overline{3}$ & $1$ & $+\nf{1}{3}$ & $+2$ & $+i$ & $+i$ & $-1$ & +1 \\
$\hat{u}^{c}_{Ri}$ & $\overline{3}$ & $1$ & $-\nf{2}{3}$ & $+1$ & $+i$ & $+i$ & $-1$ & +1 \\\hline
$\hat{L}_{Li}$ & $1$ & $2$ & $-\nf{1}{2}$ & $+2$ & $-i$ & $-i$ & $-1$ & +1 \\
$\hat{e}^{c}_{Ri}$ & $1$ & $1$ & $+1$ & $+1$ & $-i$ & $-i$ & $-1$ & +1 \\
$\hat{N}^{c}_{i}$ & $1$ & $1$ & $0$ & $0$ & $-i$ & $-i$ & $-1$ & +1 \\\hline
$\hat{H}_{d3}$ & $1$ & $2$ & $-\nf{1}{2}$ & $-3$ & $+i$ & $+i$ & +1 & +1 \\
$\hat{H}_{u3}$ & $1$ & $2$ & $+\nf{1}{2}$ & $-2$ & $+i$ & $+i$ & +1 & +1 \\
$\hat{S}_3$ & $1$ & $1$ & $0$ & $+5$ & $+i$ & $+i$ & +1 & +1 \\\hline
$\hat{H}_{d\alpha}$ & $1$ & $2$ & $-\nf{1}{2}$ & $-3$ & $+i$ & $+i$ & +1 & $e^{\f{2i\pi}{3}}$ \\
$\hat{H}_{u\alpha}$ & $1$ & $2$ & $+\nf{1}{2}$ & $-2$ & $+i$ & $+i$ & +1 & $e^{\f{2i\pi}{3}}$ \\
$\hat{S}_\alpha$ & $1$ & $1$ & $0$ & $+5$ & $+i$ & $+i$ & +1 & $e^{\f{2i\pi}{3}}$ \\\hline
$\hat{\bar{D}}_i$ & $\overline{3}$ & $1$ & $+\nf{1}{3}$ & $-3$ & $+i$ & $-i$ & +1 & $e^{\f{2i\pi}{3}}$ \\
$\hat{D}_i$ & $3$ & $1$ & $-\nf{1}{3}$ & $-2$ & $+i$ & $-i$ & +1 & $e^{\f{2i\pi}{3}}$ \\\hline
$\hat{\cc{S}}$ & 1 & 1 & 0 & $+5$ & +1 & +1 & +1 & +1 \\
$\hat{\bar{\cc{S}}}$ & 1 & 1 & 0 & $-5$ & +1 & +1 & +1 & +1 \\
$\hat{\Sigma}$ & 1 & 1 & 0 & 0 & $-i$ & $-i$ & +1 & +1 \\\hline
${\cc{W}}$ & 1 & 1 & 0 & 0 & $-i$ & $-i$ & +1 & +1 \\
${\rr{d}\theta^2}$ & 1 & 1 & 0 & 0 & $+i$ & $+i$ & +1 & +1 \\\hline
\end{tabular}
\caption{The gauge group representations and charges and
the phase changes under discrete transformations
of the superfields and superpotential of the NMSSM+.
$i$ labels all three generations whereas $\alpha$ labels the inert generations 1 and 2 only.
\label{sym}}
\end{center}
\end{table}

\subsection{The low energy NMSSM+}

The renormalisable part of the low energy effective superpotential which
respects the gauge symmetries, the $\dd{Z}_4$ $R$-symmetry,
and matter parity $\dd{Z}_2^M$ descending from the high energy theory in
equation~(\ref{NMSSM+3}) is then
\be
W^{\rr{NMSSM+}} &=& W^{\rr{E}_6\rr{SSM}} + W_3 = W_0  + W_{1,2} + W_3,\label{WWW}
\ee
where $W_0$ and $W_{1,2}$ are familiar from the E$_6$SSM, with
$W_{1,2}$ referring to either $W_1$ or $W_2$ depending on which option for the $R$-symmetry is chosen.
Once the approximate, generation-dependant $\dd{Z}_3^{HD}$ symmetry
is imposed $W_0$ approximately becomes
\be
W_0' &=& \lambda\hat{S}_3\hat{H}_{d}\hat{H}_{u}
+ \lambda_{\alpha\beta\gamma}\hat{S}_\alpha\hat{H}_{d\beta}\hat{H}_{u\gamma}
+ \kappa_{\alpha jk}\hat{S}_\alpha\hat{\bar{D}}_j\hat{D}_k
+ h^N_{3jk}\hat{H}_{u3}\hat{L}_{Lj}\hat{N}^c_k\nn\\
&& \phantom{~} + h^U_{3jk}\hat{H}_{u3}\hat{Q}_{Lj}\hat{u}_{Rk}^c
+ h^D_{3jk}\hat{H}_{d3}\hat{Q}_{Lj}\hat{d}_{Rk}^c + h^E_{3jk}\hat{H}_{d3}\hat{L}_{Lj}\hat{e}_{Rk}^c.
\ee
Note that the Yukawa couplings $h$
other than $h^N_{3jk}$, $h^U_{3jk}$, $h^D_{3jk}$, and $h^E_{3jk}$ are suppressed
and that the last four terms above correspond to $W_{\rr{Yukawa}}$ in equation~(\ref{yuk}),
coupling the active third generation of Higgs doublets to matter.
All terms in $W_{1,2}$ are also suppressed under $\dd{Z}_3^{HD}$.

The additional scale invariant term $W_3$ not present in the \essm is given by
\be
W_3 &=& \f{k_{ijk}}{3}\hat{S}_i\hat{S}_j\hat{S}_k + \f{t_{ijk}}{2}\hat{S}_i\hat{N}^c_j\hat{N}^c_k\label{w3}.
\ee
Imposing the approximate $\dd{Z}_3^{HD}$ $W_3$ approximately becomes
\be
W_3' &=& \f{k}{3}\hat{S}^3 + \f{k_{\alpha\beta\gamma}}{3}\hat{S}_\alpha\hat{S}_\beta\hat{S}_\gamma
+ \f{t_{ij}}{2}\hat{S}\hat{N}^c_i\hat{N}^c_j,\label{w3prime}
\ee
keeping only the non-suppressed couplings. 
The term $ \f{t_{ij}}{2}\hat{S}\hat{N}^c_i\hat{N}^c_j$ may be
negligible compared to other sources of right-handed neutrino masses
(see subsection~\ref{rhneu}).

The low energy NMSSM+ is therefore given approximately as 
\be
W^{\rr{NMSSM+}} &\approx & W^{\rr{E}_6\rr{SSM}}+ W'_3 \approx W'_0 + W_{1,2} + W'_3 .\label{WWWapprox}
\ee
This way of writing the low energy theory shows that the NMSSM+ can be
regarded as the \essm {\em plus}
the scale invariant cubic singlet couplings in $ W'_3$.
Note that the low energy superpotential in equation~(\ref{WWWapprox}) is 
equivalent to equation~(\ref{NMSSM+}) under the above approximations.
Thus there are two equivalent ways of looking at the NMSSM+, namely either as an extension 
of the NMSSM by the inclusion of an exotic sector, or as an extension of the \essm by the inclusion of cubic singlet
terms (with the understanding that the $U(1)_N$ is broken at high energies).

\section{Aspects of the NMSSM+}

In this section we discuss some interesting
theoretical and phenomenological aspects of the NMSSM+.
We give a commentary concerning the different symmetries
of the high energy theory and show how they result in the low energy NMSSM+ 
described in the previous section.

\subsection{Discussion of the $\dd{Z}_4$ global $R$-symmetry}

In the NMSSM+ instead of $\dd{Z}_2^{qq}$ or $\dd{Z}_2^{lq}$ we impose either
$\dd{Z}_4^{qq}$ or $\dd{Z}_4^{lq}$. These are $R$-symmetries under which the 
superfields and indeed the superpotential itself transform
as follows in each case:
\be
&\dd{Z}_4^{qq}&\nn\\
(\hat{S},\hat{H}_d,\hat{H}_u,\hat{\bar{D}},\hat{D},\hat{Q}_L,\hat{u}_R^c,\hat{d}_R^c)
&\rightarrow& e^{\f{i\pi}{2}}(\hat{S},\hat{H}_d,\hat{H}_u,\hat{\bar{D}},\hat{D},\hat{Q}_L,\hat{u}_R^c,\hat{d}_R^c),\nn\\
(\hat{L}_L,\hat{e}_R^c,\hat{N}^c) &\rightarrow& e^{\f{3i\pi}{2}}(\hat{L}_L,\hat{e}_R^c,\hat{N}^c),\nn\\
(\hat{\cc{S}},\hat{\bar{\cc{S}}}) &\rightarrow& (\hat{\cc{S}},\hat{\bar{\cc{S}}}),\nn\\
\hat{\Sigma} &\rightarrow& e^{\f{3i\pi}{2}}\hat{\Sigma},\nn\\
\cc{W} &\rightarrow& e^{\f{3i\pi}{2}}\cc{W}\\[2mm]
\Big[\Rightarrow\quad\rr{d}\theta^2 &\rightarrow& e^{\f{-3i\pi}{2}}\rr{d}\theta^2
= e^{\f{i\pi}{2}}\rr{d}\theta^2\Big].\nn
\ee
\be
&\dd{Z}_4^{lq}&\nn\\
(\hat{S},\hat{H}_d,\hat{H}_u,\hat{Q}_L,\hat{u}_R^c,\hat{d}_R^c)
&\rightarrow& e^{\f{i\pi}{2}}(\hat{S},\hat{H}_d,\hat{H}_u,\hat{Q}_L,\hat{u}_R^c,\hat{d}_R^c),\nn\\
(\hat{\bar{D}},\hat{D},\hat{L}_L,\hat{e}_R^c,\hat{N}^c) &\rightarrow& e^{\f{3i\pi}{2}}(\hat{\bar{D}},\hat{D},\hat{L}_L,\hat{e}_R^c,\hat{N}^c),\nn\\
(\hat{\cc{S}},\hat{\bar{\cc{S}}}) &\rightarrow& (\hat{\cc{S}},\hat{\bar{\cc{S}}}),\nn\\
\hat{\Sigma} &\rightarrow& e^{\f{3i\pi}{2}}\hat{\Sigma},\nn\\
\cc{W} &\rightarrow& e^{\f{3i\pi}{2}}\cc{W}\\[2mm]
\Big[\Rightarrow\quad\rr{d}\theta^2 &\rightarrow& e^{\f{-3i\pi}{2}}\rr{d}\theta^2
= e^{\f{i\pi}{2}}\rr{d}\theta^2\Big].\nn
\ee

The renormalisable high energy superpotential with $\dd{Z}_4^{qq,lq}$ is then
\be
\label{ren}
\cc{W}^{\rr{NMSSM+}}_{\rr{ren}} &=& W_0 + W_{1,2} + \hat{\Sigma}\left(l\hat{\cc{S}}\hat{\bar{\cc{S}}} - M_\Sigma^2\right),
\ee
which includes only the renormalisable terms of $\cc{W}^{\rr{NMSSM+}}$ in equation~(\ref{NMSSM+3}).
We have thus found $R$-symmetries that allow the usual trilinear superpotential terms of the
E$_6$SSM, avoid rapid proton decay by forbidding either $W_1$ or $W_2$, and also
distinguish between the conventional SM-singlets $S_i$ and the new $\cc{S}$
in a way that allows for
the $\Sigma$ scale and the EWSB scale to be naturally separated
even though $S_i$ and $\cc{S}$ share the same
gauge charges. The further consequences of these $R$-symmetries are discussed below.

\subsection{High energy $U(1)_N$ gauge symmetry breaking in the NMSSM+}

At the $\Sigma$ scale there is no radiative EWSB.
The high energy scalar potential relevant for finding the
VEVs generated at the $\Sigma$ scale is obtained from 
the renormalisable terms in $W^{\cc{S}}$ in equation~(\ref{wS}),
together with $D$-terms and soft mass terms,
\be
\cc{V} &=& |l\cc{S}\bar{\cc{S}} - M_\Sigma^2|^2\nn\\
&&\phantom{~} + l^2|\Sigma|^2\left(|\cc{S}|^2 + |\bar{\cc{S}}|^2\right) + \f{D_N^2}{2}\nn\\
&&\phantom{~} + m_{\cc{S}}^2|\cc{S}|^2 + m_{\bar{\cc{S}}}^2|\bar{\cc{S}}|^2
+ m_{\Sigma}^2|\Sigma|^2\nn\\
&&\phantom{~} + \mbox{terms with zero VEV},
\ee
where $m_{(\cc{S},\bar{\cc{S}},\Sigma)}^2$ are soft supersymmetry breaking
masses-squared and
\be
D_N &=& g_1^\prime\left(Q^N_S|\cc{S}|^2-Q^N_S|\bar{\cc{S}}|^2+\sum_aQ^N_a|\varphi_a|^2\right)
\ee
is the $U(1)_N$ $D$-term, with $\varphi_a$ and $Q^N_a$ the other scalars
(the E$_6$SSM-like scalars that survive to low energy) and their $U(1)_N$ charges.
Minimising the scalar potential with respect to $\cc{S}$, $\bar{\cc{S}}$, and
$\Sigma$ gives
\be
|\langle\cc{S}\rangle|^2 &=& \f{M_\Sigma^2}{l} + \cc{O}(m_{\cc{S}}^2,m_{\bar{\cc{S}}}^2),\\
|\langle\bar{\cc{S}}\rangle|^2 &=& \f{M_\Sigma^2}{l} + \cc{O}(m_{\cc{S}}^2,m_{\bar{\cc{S}}}^2),\\
\langle\Sigma\rangle &=& 0,\\
\langle D_N\rangle &=& \f{m_{\bar{\cc{S}}}^2 - m_{\cc{S}}^2}{2g'_1Q^N_S}.
\ee
The first term in the potential sets the scale for
the $\cc{S}$ and $\bar{\cc{S}}$
VEVs to $M_\Sigma$ and the form of the $D$-term
requires them to be equal to be each other up to possible corrections
of order the soft supersymmetry breaking scale. This is why we include the
\textit{pair} of extra fields with opposite $U(1)_N$ charges,
so that the minimisation conditions lead to $U(1)_N$ breaking
that is approximately $D$-flat.

The VEV for the $U(1)_N$ $D$-term, proportional to $m_{\bar{\cc{S}}}^2 - m_{\cc{S}}^2$,
may be small, or even zero, depending on the nature of supersymmetry breaking
and its mediation to the visible sector. Below the $\Sigma$ scale, when the
fields that acquire masses are integrated out, any soft scale $\langle D_N\rangle$ will
just lead to corrections for the effective soft supersymmetry breaking masses-squared
for the other scalars
\be
\Delta m_a^2 &=& \f{Q^N_a}{Q^N_S}(m_{\bar{\cc{S}}}^2 - m_{\cc{S}}^2),
\ee
as discussed in \cite{Kolda:1995iw}.

\subsection{How the NMSSM+ solves the domain wall problem}
\label{domain}

If $M_\Sigma$ is not too far below the unification scale $M_X$ scale then an effective
NMSSM-like $\hat{S}^3$ term is generated by 
the first non-renormalisable term in $W^{\cc{S}}$ in equation~(\ref{wS}),
\be
\f{b_{333}}{M_X^3}\hat{S}^3\langle\hat{\bar{\cc{S}}}^3\rangle,
\ee
with a coefficient $k/3$ not too far from unity. This will break the otherwise present global
$U(1)$ symmetry of the effective low energy theory below the $\Sigma$ scale
down to a global $\dd{Z}_3$~\footnote{ This $U(1)$ symmetry is familiar in the
NMSSM in the limit of vanishing $S^3$ term, and the same symmetry 
in the conventional E$_6$SSM is gauged to become the $U(1)_N$
symmetry. Here the $U(1)_N$ gauge symmetry is spontaneously broken 
by the high scale $\bar{\cc{S}}$ VEV, resulting at the low energy scale in an
effective $\hat{S}^3$ term in the superpotential.}.


\subsubsection{Domain wall destabilisation}

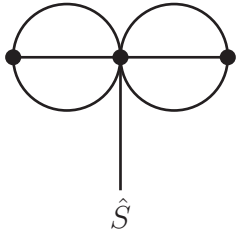
\begin{figure}
\begin{center}
\begin{picture}(100,80)(-50,-10)
\SetWidth{1}
\Line(0,0)(0,50)
\Text(0,-3)[t]{$\hat{S}$}
\Vertex(0,50){3}
\CArc(20,50)(20,0,360)
\CArc(-20,50)(20,0,360)
\Line(-40,50)(40,50)
\Vertex(-40,50){3}
\Vertex(40,50){3}
\end{picture}
\end{center}
\caption{The form of SM-singlet tadpoles caused by the existence of dimension 7
terms in the superpotential as well as dimension 3 terms.\label{tadpole}}
\end{figure}

The discrete $R$-symmetry of the NMSSM+ also allows the non-renormalisable terms
in square brackets in equation~(\ref{wS}),
\be
\f{c}{M_X^{11}}\hat{S}^7\hat{\bar{\cc{S}}}^7
+ \f{c'}{M_X^8}\hat{S}^5\hat{H}_d.\hat{H}_u\hat{\bar{\cc{S}}}^4
+ \f{c''}{M_X^5}\hat{S}^3(\hat{H}_d.\hat{H}_u)^2\hat{\bar{\cc{S}}}
\ee
to appear in the high energy superpotential.
When the high scale $\bar{\cc{S}}$ VEV is inserted 
these terms generate the effective dimension 7 terms
$\hat{S}^7$, $\hat{S}^5(\hat{H}_d.\hat{H}_u)$, and $\hat{S}^3(\hat{H}_d.\hat{H}_u)^2$,
each suppressed by 4 powers of $M_X$, in the low energy superpotential.
Through 4-loop tadpoles of the form in Figure~\ref{tadpole} these dimension 7
terms and the dimension 3 terms will generate linear terms in the potential of order~\cite{Abel:1996cr}
\be
\f{1}{(16\pi^2)^4}m_{\rr{soft}}^3(S+S^*),
\ee
breaking the accidental global $\dd{Z}_3$ and destabilising its cosmological
domain walls.

\subsubsection{Forbidden non-renormalisable operators and lack of divergences}

It is important that the above terms are allowed by the $R$-symmetry.
As well as these terms containing 7 low energy (conventional E$_6$SSM)
superfields, the $R$-symmetry of the NMSSM+ also allows terms containing 5 low
energy superfields (where at least one superfield is a lepton or exotic coloured
superfield). Non-renormalisable terms containing just 3 low energy superfields
are discussed in the following subsection along with K\"ahler potential terms
containing just two low energy superfields.

Superpotential terms containing
even numbers of low energy superfields and K\"ahler potential terms containing
odd numbers of low energy superfields are completely forbidden by the
$R$-symmetry. This is important because such terms would induce effective terms
in the low energy superpotential that in the NMSSM have been shown to lead to
dangerously divergent tadpoles~\cite{Abel:1996cr}.

\subsection{Matter parity violating operators}
\label{RPV}

For completeness we list the following matter parity violating
terms which are allowed by the gauge symmetry ($\forall\;i,j,k$)
but are forbidden by either $\dd{Z}_4^{qq}$ or $\dd{Z}_4^{lq}$ in the superpotential:
\be
&\hat{L}_{Li}\hat{L}_{Lj}\hat{e}_{Rk}^c\f{\hat{\bar{\cc{S}}}}{M_X},\quad
\hat{Q}_{Li}\hat{L}_{Lj}\hat{d}_{Rk}^c\f{\hat{\bar{\cc{S}}}}{M_X},\quad
\hat{H}_{di}\hat{H}_{dj}\hat{e}_{Rk}^c\f{\hat{\cc{S}}}{M_X},&\nn\\
&\hat{S}_i\hat{H}_{uj}\hat{L}_{Lk}\f{\hat{\bar{\cc{S}}}}{M_X},\quad
\hat{N}_i^c\hat{H}_{dj}\hat{H}_{uk}\f{\hat{\cc{S}}}{M_X};&
\ee
and in the K\"ahler potential:
\be
&\hat{L}_{Li}\hat{H}_{dj}^\dagger\f{\hat{\bar{\cc{S}}}}{M_X} + \mbox{c.c.},\quad
\hat{L}_{Li}\hat{H}_{dj}^\dagger\f{\hat{\cc{S}}^\dagger}{M_X} + \mbox{c.c.},&\nn\\
&\hat{N}^c_i\hat{S}_j^\dagger\f{\hat{\cc{S}}}{M_X} + \mbox{c.c.},\quad
\hat{N}^c_i\hat{S}_j^\dagger\f{\hat{\bar{\cc{S}}}^\dagger}{M_X} + \mbox{c.c.}.&
\ee
Since we assume either $\dd{Z}_4^{qq}$ or $\dd{Z}_4^{lq}$, none of the above terms are present.

On the other hand, the following matter parity violating terms appearing in
$\Delta W_1$ in equation~(\ref{delw1}) are allowed by the gauge symmetry and
$\dd{Z}_4^{qq}$, but are forbidden by $\dd{Z}_4^{lq}$:
\be
\hat{S}_i\hat{D}_j\hat{d}_{Rk}^c\f{\bar{\hat{\cc{S}}}}{M_X},\quad
\hat{H}_{di}\hat{\bar{D}}_j\hat{Q}_{Lk}\f{\hat{\bar{\cc{S}}}}{M_X};
\ee
and the following matter parity violating terms appearing in $\Delta W_0$ in equation~(\ref{delw0}) are allowed by the gauge symmetry and
both $\dd{Z}_4^{qq}$ and $\dd{Z}_4^{lq}$:
\be
\hat{u}_{Ri}^c\hat{d}_{Rj}^c\hat{d}_{Rk}^c\f{\hat{\bar{\cc{S}}}}{M_X},\quad
\hat{\bar{D}}_i\hat{\bar{D}}_j\hat{u}_{Rk}^c\f{\hat{\cc{S}}}{M_X}.
\ee
The following matter parity violating K\"ahler potential terms $\Delta K_1$
are also allowed by the gauge symmetry and
$\dd{Z}_4^{qq}$, but forbidden by $\dd{Z}_4^{lq}$:
\be
\Delta K_1 &=& \f{r_{ij}^{dD}}{M_X}\hat{d}^c_{Ri}\hat{\bar{D}}_j^\dagger\hat{\bar{\cc{S}}}
+ \f{r_{ij}^{dD\prime}}{M_X}\hat{d}^c_{Ri}\hat{\bar{D}}_j^\dagger\hat{\cc{S}}^\dagger + \mbox{c.c.}.
\ee

These terms are all forbidden by imposing the matter parity $\dd{Z}_2^M$, but we list them
in case one wishes to relax $\dd{Z}_2^M$. We give the two cases, for the two $R$-parity choices, below.

\subsubsection{$\dd{Z}_4^{qq}$ without $\dd{Z}_2^M$}

In this case $\Delta W_0$, $\Delta W_1$, and $\Delta K_1$ are all allowed and
the low energy superpotential is given by
\be
W^{\rr{NMSSM+}} &=& W_0 + W_1 + W_3 + \Delta W_0 + \Delta W_1.
\ee
The $\hat{D}$ and $\hat{\bar{D}}$ are interpreted as anti-diquarks and diquarks
respectively so that $W_0 + W_1$ respects $B$ and $L$ (baryon and lepton number)
conservation. In this case
$\Delta W_0$, $\Delta W_1$, and $\Delta K_1$ break $B$, but respect $L$ 
\footnote{RH neutrinos violate lepton number conservation
via the term in $W_3$ and via any large intermediate scale Majorana mass term
(see subsection~\ref{rhneu}), both
leading to $\Delta L = \pm 2$ effects. This does not affect the arguments of
this subsection}.
It is clear that since $L$ is conserved, proton decay with one lepton in the
final state is forbidden.
More generally, since \emph{every one} of the matter parity violating
couplings, coming from $\Delta W_0$, $\Delta W_1$, and $\Delta K_1$, change
$B$ by exactly one, $\Delta B = -1$
proton decay is forbidden as long as having any supersymmetric particles in the final
state is kinematically not allowed.

Rapid proton decay is therefore avoided, but the LSP would be allowed to
decay; for example, the effective superpotential
term $\hat{u}_{Ri}^c\hat{d}_{Rj}^c\hat{d}_{Rk}^c$
allows the decays
$\ii{\chi}^0 \rightarrow p^\pm K^\mp$ if
the neutralino $\ii{\chi}^0$ contains some admixture of the bino ($\tilde{B}$),
the superpartner of the Abelian hypercharge gauge boson.

While we observe that rapid proton decay is avoided in this scenario,
certainly as long as the lightest supersymmetric particle is heavier than the proton,
$\Delta B =\pm 2$ effects~\cite{Zwirner:1984is}, such as
$n$-$\bar{n}$ oscillations~\cite{Abe:2011ky} and dinucleon decay $p^+p^+ \rightarrow K^+K^+$,
would also need to be considered.


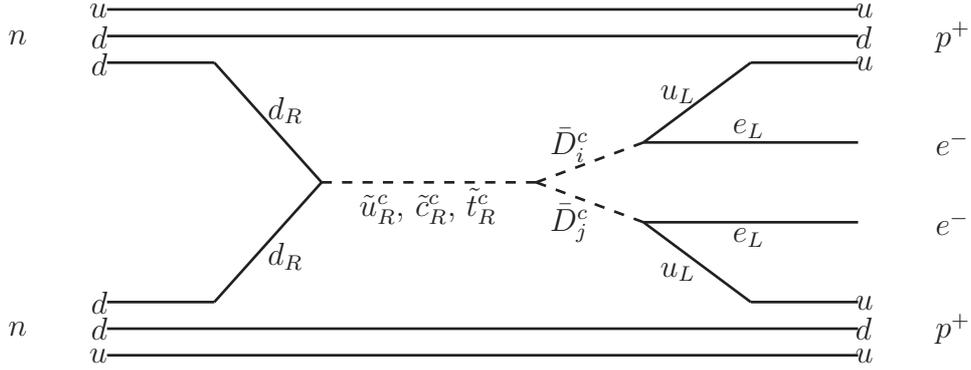
\begin{figure}
\begin{center}
\begin{picture}(280,130)(-40,0)
\SetWidth{1}
\Line(-40,0)(240,0)
\Line(-40,10)(240,10)
\Line(-40,20)(0,20)
\Line(200,20)(240,20)
\Line(0,20)(40,65)
\DashLine(40,65)(120,65){4}
\DashLine(120,65)(160,50){4}
\Line(160,50)(200,20)
\Line(160,50)(240,50)
\Line(-40,130)(240,130)
\Line(-40,120)(240,120)
\Line(-40,110)(0,110)
\Line(200,110)(240,110)
\Line(0,110)(40,65)
\DashLine(120,65)(160,80){4}
\Line(160,80)(200,110)
\Line(160,80)(240,80)
\Text(-40,0)[r]{$u$}
\Text(-40,10)[r]{$d$}
\Text(-40,20)[r]{$d$}
\Text(-40,110)[r]{$d$}
\Text(-40,120)[r]{$d$}
\Text(-40,130)[r]{$u$}
\Text(240,0)[l]{$u$}
\Text(240,10)[l]{$d$}
\Text(240,20)[l]{$u$}
\Text(240,110)[l]{$u$}
\Text(240,120)[l]{$d$}
\Text(240,130)[l]{$u$}
\Text(140,57)[tr]{$\bar{D}^c_j$}
\Text(140,73)[br]{$\bar{D}^c_i$}
\Text(180,35)[tr]{$u_L$}
\Text(180,95)[br]{$u_L$}
\Text(200,48)[t]{$e_L$}
\Text(200,82)[b]{$e_L$}
\Text(270,10)[l]{$p^+$}
\Text(270,50)[l]{$e^-$}
\Text(-70,10)[r]{$n$}
\Text(270,120)[l]{$p^+$}
\Text(270,80)[l]{$e^-$}
\Text(-70,120)[r]{$n$}
\Text(20,42.5)[tl]{$d_R$}
\Text(20,87.5)[bl]{$d_R$}
\Text(80,63)[t]{$\tilde{u}_R^c$, $\tilde{c}_R^c$, $\tilde{t}_R^c$}
\end{picture}
\end{center}
\caption{A Feynman diagram for neutrinoless double beta decay in the theory where
$\dd{Z}_4^{lq}$ is imposed without matter parity $\dd{Z}_2^M$.
The two couplings involving squarks are from the two matter parity violating terms
in $\Delta W_0$. Here they are used together to change lepton number by two
while preserving baryon number.
The first vertex is at least CKM suppressed since
$r^{udd}_{ijk}$ from equation~(\ref{delw0}) has to be antisymmetric in its
last two indices ($jk$) in the gauge basis (due to $SU(3)$ gauge symmetry)
and therefore cannot couple two down quark interaction states together.
The second vertex is from the soft $A$ term associated with $r^{DDu}_{ijk}$.
\label{0nbb}}
\end{figure}

\subsubsection{$\dd{Z}_4^{lq}$ without $\dd{Z}_2^M$}

In this case $\Delta W_0$, but neither $\Delta W_1$ nor $\Delta K_1$, is allowed and
\be
W^{\rr{NMSSM+}} &=& W_0 + W_2 + W_3 + \Delta W_0,
\ee
where
\be
W_2 &=& g^N_{ijk}\hat{N}^c_i\hat{D}_j\hat{d}_{Rk}^c + g^E_{ijk}\hat{D}_i\hat{u}_{Rj}^c\hat{e}_{Rk}^c +
g^D_{ijk}\hat{\bar{D}}_i\hat{Q}_{Lj}\hat{L}_{Lk}.
\ee
Without $\Delta W_0$, just looking at $W_0 + W_2$,
$\hat{D}$ and $\hat{\bar{D}}$ would be interpreted as leptoquarks
and anti-leptoquarks respectively, implying $B$ and $L$ conservation.
The presence of $\Delta W_0$ introduces the terms
$\hat{\bar{D}}_i\hat{\bar{D}}_j\hat{u}_{Rk}^c$ and $\hat{u}_{Ri}^c\hat{d}_{Rj}^c\hat{d}_{Rk}^c$.
With the above interpretation these terms have $(\Delta B,\Delta L) = (-1,-2)$
and~$(-1,0)$ respectively.
Since $\Delta L =\pm 1$ terms are absent, proton decay with one lepton in the
final state is again forbidden. More generally, in this scenario too, every one
of the matter parity violating couplings, coming from $\Delta W_0$, change
$B$ by exactly one, meaning that once again $\Delta B = -1$
proton decay is forbidden without supersymmetric particles in the final state.

Once again $\Delta B =\pm 2$ effects would have to be considered, but now so too
would $\Delta L =\pm 2$ effects.
For instance, both of the matter parity violating terms from $\Delta W_0$
can be combined with $\hat{\bar{D}}_i\hat{Q}_{Lj}\hat{L}_{Lk}$ from $W_2$
to give contributions
to neutrinoless double beta decay, as shown in Figure~\ref{0nbb},
different from those considered in~\cite{Deppisch:2012nb}.

\subsection{Harmless non-renormalisable terms}
For each of the superpotential and K\"ahler potential terms that we have mentioned so far that are invariant
with respect to the gauge symmetry and the $\dd{Z}_4$ global $R$-symmetry and
optionally matter parity there is also the same term
multiplied by any arbitrary power of $(\cc{S}\bar{\cc{S}})/M^2_X$ with some coefficient.
Assuming that 
\be
\langle\cc{S}\bar{\cc{S}}\rangle &\approx& \f{M_\Sigma^2}{l}\nn
\ee
is somewhat less than $M_X^2$ these higher order terms
will be sub-dominant in the perturbative expansion of the low energy theory,
and represent harmless corrections to the leading terms.

\subsection{The suppression of flavour changing neutral currents and the generation of
exotic D-fermion masses}

We now motivate the approximate $\dd{Z}_3^{HD}$ given in Table~\ref{sym} as
a way to remove flavour changing neutral currents and allow the generation of D-fermion masses
from inert SM-singlet VEVs that can naturally be at a scale slightly larger than
the EWSB scale.

Let us examine the Yukawa couplings present in the low energy NMSSM+ potential
given in equation~(\ref{WWW})
\be
W^{\rr{NMSSM+}} &=& W_0  + W_{1,2} + W_3.
\ee
$W_0$ contains Yukawa couplings of all three generations of Higgs doublets
to matter as well as
\be
\lambda_{ijk}\hat{S}_i\hat{H}_{dj}\hat{H}_{uk} + \kappa_{ijk}\hat{S}_i\hat{\bar{D}}_j\hat{D}_k.\nn
\ee
Note that SM-singlet VEVs are the only way for the exotic D-fermions to
acquire mass (the scalar components also acquire mass through soft supersymmetry breaking terms).

In the NMSSM+ we propose that the inert SM-singlets $S_\alpha$ acquire VEVs and that
these VEVs are responsible for the D-fermion masses, rather than the active
SM-singlet VEV $s = s_3$, where we define $\alpha,\beta,\gamma\in\{1,2\}$ and
$s_i = \sqrt{2}\langle S_i\rangle$.
It is important that the $\hat{\bar{D}}$ and $\hat{D}$ receive mass from 
different singlets to those responsible for the $\mu$ term since otherwise
their mass limits would lead to larger values of $\mu$ and hence to fine-tuning.
By contrast, the VEVs of the other (inert) SM-singlets ${S}_\alpha$
may be slightly larger than the EWSB scale, providing large exotic
masses, without necessarily introducing fine-tuning. 
This would not be advantageous in the conventional E$_6$SSM, since
any and all of the three possible SM-singlet VEVs would contribute
to $\langle D_N\rangle$ and any large SM-singlet VEV would therefore lead
to fine-tuning, as explained in subsection~\ref{sub:confine}.

In the conventional E$_6$SSM the approximate flavour symmetry $\dd{Z}_2^H$,
under which only the third (active) generation of Higgs doublets and SM-singlets are even,
is applied to suppress the Yukawa couplings of the inert generations of Higgs doublets to matter,
suppressing flavour changing neutral currents and explaining why only the active generation
radiatively acquires VEVs. In the NMSSM+ something different is required to do this job
since $\dd{Z}_2^H$ would suppress couplings of the form $\kappa_{\alpha ij}$.
In the NMSSM+ these couplings should be large both to explain how
large inert singlet VEVs could be radiatively induced and because
these couplings then appear in the masses induced for the D-fermions by those inert singlet VEVs.
Furthermore, in order to keep a slight hierarchy $s_\alpha \gg s_3$ natural
terms coupling the EWSB scale VEVs $s_3$, $v_d$, and $v_u$ to the larger VEVs
$s_2$ and $s_1$ should not be very large. We therefore also require an approximate
flavour symmetry that suppresses trilinear SM-singlet couplings
$k_{ijk}$ from $W_3$ other than $k = k_{333}$ and those of the form $k_{\alpha\beta\gamma}$,
as well as various $\lambda_{ijk}$ couplings, namely those of the forms
$\lambda_{\alpha 3i}$ and $\lambda_{\alpha i3}$~\footnote{We note that
these are the ``f- and z-couplings'' that are required
to be exactly zero in the EZSSM dark matter scenario~\cite{Hall:2011zq}.
In this version of the model a discrete symmetry $\dd{Z}_2^S$
is imposed under which only $\hat{S}_\alpha$ are odd. Unfortunately this symmetry would also forbid
$\kappa_{\alpha ij}$ which should be large in the NMSSM+
for the reasons described.}.

In the NMSSM+, instead of $\dd{Z}_2^H$, we have an appoximate symmetry
$\dd{Z}_3^{HD}$ under which
\be
&\dd{Z}_3^{HD}&\nn\\
(\hat{S}_\alpha,\hat{H}_{d\alpha},\hat{H}_{u\alpha},\hat{\bar{D}}_i,\hat{D}_i)
&\rightarrow& e^{\f{2i\pi}{3}}(\hat{S}_\alpha,\hat{H}_{d\alpha},\hat{H}_{u\alpha},\hat{\bar{D}}_i,\hat{D}_i)
\ee
and no other superfields transform.
This then leads to the approximate NMSSM+ superpotential given in equation~(\ref{NMSSM+}).
Like $\dd{Z}_2^H$, this approximate symmetry suppresses Yukawa couplings of the inert generations of Higgs doublets
to matter and all couplings in $W_{1,2}$, suppressing flavour changing neutral currents.
It also suppresses all $k_{ijk}$ couplings other than $k = k_{333}$
and those of the form $k_{\alpha\beta\gamma}$;
all $\lambda_{ijk}$ couplings other than $\lambda = \lambda_{333}$
and those of the form $\lambda_{\alpha\beta\gamma}$;
and all $\kappa_{ijk}$ couplings other than those of the form $\kappa_{\alpha jk}$.

We emphasise that the $\dd{Z}_3^{HD}$ symmetry should not be exact.
An exact $\dd{Z}_3^{HD}$ symmetry would exactly forbid both $W_1$ and $W_2$,
meaning that the exotic coloured $\hat{D}$ and $\hat{\bar{D}}$ particles would not
be able to decay (the same reason why $\dd{Z}_2^H$ cannot be exact in the E$_6$SSM.)
Furthermore, since the inert SM-singlets transform under the symmetry the exact symmetry
would be spontaneously broken by the inert SM-singlet VEVs and this would lead to
cosmological domain walls. This symmetry is therefore regarded as an approximate
flavour symmetry, on the same footing as the approximate $\dd{Z}_2^{H}$ in the E$_6$SSM.
Since the symmetry is not exact small effective linear terms for the inert SM-singlets, which break
$\dd{Z}_3^{HD}$, will be generated for the inert SM-singlets in the same
way that they are generated for the active SM-singlet, via tadpole diagrams of the form in
Figure~\ref{tadpole} in subsection~\ref{domain}.

\subsection{Right-handed neutrino masses}
\label{rhneu}

It is important to note that the $\dd{Z}_4$ $R$-symmetry forbids Planck scale Majorana
RH neutrino masses which are otherwise allowed since RH neutrinos are complete gauge singlets.

Once the $\dd{Z}_3^{HD}$ symmetry from the previous subsection has been applied
$W_3$ from equation~(\ref{w3}) approximately becomes
\be
W_3' &=& \f{k}{3}\hat{S}^3 + \f{k_{\alpha\beta\gamma}}{3}\hat{S}_\alpha\hat{S}_\beta\hat{S}_\gamma
+ \f{t_{ij}}{2}\hat{S}\hat{N}^c_i\hat{N}^c_j
\ee
as in equation~(\ref{w3prime}).
A Majorana RH neutrino mass is therefore generated by the active EWSB scale SM-singlet VEV $\langle S\rangle = s_3/\sqrt{2}$.

However, some extra mechanism would have to be responsible for generating
an intermediate scale Majorana mass as needed for a type-I see-saw mechanism.
This could be achieved, for example, by a complete
gauge singlet $\hat{S}_N$ that transforms under the $R$-symmetry as
\be
\hat{S}_N &\rightarrow& e^{\f{i\pi}{2}}\hat{S}_N = +i\hat{S}_N
\ee
and couples to the RH neutrinos via the superpotential term
\be
\f{e_{ij}}{2}\hat{S}_N\hat{N}^c_i\hat{N}^c_j,
\ee
with $S_N$ acquiring an appropriate intermediate scale VEV.

\subsection{Grand unification}

The situation with respect to grand unification is similar to that in the E$_6$SSM.
The extra $SU(2)$ doublets $H'$ and $\bar{H}'$ mentioned in
the Introduction, which are included in the
\essm \cite{King:2007uj} for gauge coupling unification, may
be included with a mass of order 10~TeV.

Alternatively in the NMSSM+ the $H'$ and $\bar{H}'$ may be omitted.
Without these fields and with the $E_6$ assumed broken directly to
$G_{\rr{SM}} \otimes U(1)_N$ at the GUT scale,
the renormalisation group equations do not actually allow the
gauge couplings to unify below the Planck scale.
However, since in the NMSSM+ we already have an intermediate
scale $M_\Sigma$ slightly below the assumed grand unification scale, 
one could assume that something like in the Minimal E$_6$SSM~\cite{Howl:2007zi} happens,
where $E_6$ is not broken directly to $G_{\rr{SM}} \otimes U(1)_N$ and where the running is
modified above some intermediate scale.

Whichever scenario is chosen, the fact that the matter content consists of complete $27$ representations of $E_6$
(plus $\cc{S}$, $\cc{\bar{S}}$, $\Sigma$, and possibly $H'$ and $\bar{H}'$) ensures anomaly cancellation from the GUT scale
to the $\Sigma$ scale where $U(1)_N$ is broken.

\section{Fine-Tuning}
In this section we give a qualitative discussion of the tree-level fine-tuning which is present in the E$_6$SSM,
and then show how the NMSSM+ leads to a dramatic improvement in tree-level fine-tuning.

\subsection{Tree-level fine-tuning in the conventional E$_6$SSM}
\label{sub:confine}

The E$_6$SSM active scalar potential relevant for EWSB may be written \cite{King:2005jy}
\be
{V} &=& \lambda^2|S|^2\left(|H_{d}|^2+|H_{u}|^2\right) + \lambda^2|H_{d}.H_{u}|^2\nn\\
&&\phantom{~} + \f{g_1^2}{8}\left(|H_{d}|^2-|H_{u}|^2\right)^2 + \f{g_2^2}{8}\left(H_{d}^\dagger\sigma^aH_{d}-H_{u}^\dagger\sigma^aH_{u}\right)^2
+ \f{D_N^2}{2}\nn\\
&&\phantom{~} + m_d^2|H_d|^2 + m_u^2|H_u|^2 + m_S^2|S|^2 + \left[\lambda A_\lambda SH_dH_u + \mbox{c.c.}\right],
\ee
where
\be
D_N &=& g_1^\prime\left(Q_d|H_d|^2+Q_u|H_u|^2+Q_S|S|^2+\mbox{terms with zero VEV}\right)
\ee
is the $U(1)_N$ $D$-term.

The Higgs tree-level minimisation conditions in this model are \cite{King:2005jy}
\beq \ba{rcl}
\frac{\partial {V}}{\partial s}&=&m_{S}^2 s-\frac{\lambda
A_{\lambda}}{\sqrt{2}}v_1v_2+\frac{\lambda^2}{2}(v_d^2+v_u^2)s\\[2mm]
&&\phantom{~}+\frac{g^{'2}_1}{2}\biggl(Q_dv_d^2+Q_uv_u^2+Q_S
s^2\biggr)Q_S s=0\,,
\\[2mm] \frac{\partial {V}}{\partial v_d}&=&
m_d^2v_d-\frac{\lambda A_{\lambda}}{\sqrt{2}}s v_u
+\frac{\lambda^2}{2}(v_u^2+s^2)v_d+\frac{\bar{g}^2}{8}\biggl(v_d^2-v_u^2\biggr)v_d\\[2mm]
&&\phantom{~}+\frac{g^{'2}_1}{2}\biggl(Q_dv_d^2+Q_uv_u^2+Q_Ss^2\biggr)Q_d
v_d=0\,,\\[2mm] 
\label{42}
\frac{\partial {V}}{\partial v_u}&=& m_u^2v_u-\frac{\lambda A_{\lambda}}{\sqrt{2}}s v_d+
\frac{\lambda^2}{2}(v_d^2+s^2)v_u+\frac{\bar{g}^2}{8}\biggl(v_u^2-v_d^2\biggr)v_u\\[2mm]
&&\phantom{~}+\frac{g^{'2}_1}{2}\biggl(Q_dv_d^2+Q_uv_u^2+Q_Ss^2\biggr)Q_u v_u=0\,,
\ea
\eeq
where $\bar{g}=\sqrt{g_2^2+g'^2}$, $s=\sqrt{2}\langle S\rangle$, and $v_{d,u} = \sqrt{2}\langle H_{d,u}\rangle$.

We begin by dropping factors of order unity and making the approximations $s\gg v_d , v_u$ and $\lambda \ll \bar{g}$,
and writing $v\sim v_d \sim v_u$, and $m_d^2\sim m_u^2 \sim m^2$. These equations may then be combined to yield,
\beq
M_Z^2\sim \mu A_{\lambda} -m^2 + \mu^2 -M_{Z'}^2,
\label{approxmincond}
\eeq
where $\mu \sim \lambda s$ and $M_{Z'}^2 \approx -2m_{S}^2$.
More accurate minimisation conditions will be derived below, keeping all the factors of order unity.
However equation~(\ref{approxmincond}) is sufficient to show that in order to avoid {\it tree-level} fine-tuning 
we need to keep both $\mu $ and $M_{Z'}$ as close to the electroweak scale as possible.
Clearly the recent experimental limit of $M_{Z'}>2$ TeV \cite{Chatrchyan:2012it}
leads to significant fine-tuning.
We emphasise that the appearance of $M_{Z'}$ in the tree-level minimisation condition is characteristic
of all $Z'$ models where the usual Higgs doublets carry $U(1)'$ charges (e.g. it applies to all $E_6$ models
but not the $U(1)_{B-L}$ model).

For the more accurate conditions we instead minimise with respect to
$s^2$ and $v^2=v_d^2 + v_u^2$.
Classically minimising with respect to $s^2$ yields
\be
\frac{\partial{{V}}}{\partial(s^2)} &=& 2g_1^{\prime 2}Q_S^2s^2
+4\lambda^2\left(v_d^2+v_u^2\right) + 2m_S^2 + 4g_1^{\prime 2}Q_S\left(Q_dv_d^2+Q_uv_u^2\right)
+ 2\sqrt{2}\lambda A_\lambda\f{v_dv_u}{s} = 0.\nn
\ee
The limit on $M_{Z'}^2 \approx 2g_1^{\prime 2}Q_S^2s^2$ implies that
$s^2 \gg v^2$ and this
then requires a large negative $m_S^2$. The equation approximately becomes
\be
g_1^{\prime 2}Q_S^2s^2 + m_S^2 \approx 0
&\Rightarrow& M_{Z'}^2 \approx -2m_{S}^2.
\ee
Classically minimising with respect to $v^2$ yields
\be
\frac{\partial{{V}}}{\partial(v^2)} &=& \bar{g}^2(c^2_\beta - s^2_\beta)^2v^2 + 4g_1^{\prime 2}\left(Q_dc^2_\beta + Q_us^2_\beta\right)^2v^2 + 8\lambda^2s^2_\beta c^2_\beta v^2\nn\\
&&\phantom{~} + 4\lambda^2s^2 + 2m_d^2c^2_\beta + 2m_u^2s^2_\beta\nn\\
&&\phantom{~} + 4g_1^{\prime 2}Q_Ss^2\left(Q_dc^2_\beta + Q_us^2_\beta\right) + 4\sqrt{2}\lambda A_\lambda sc_\beta s_\beta = 0.
\ee
In order to satisfy this condition with $s^2 \gg v^2$ fine-tuning is required to occur
between the $s^2$ terms, namely $\lambda^2s^2$ and
$g_1^{\prime 2}Q_Ss^2\left(Q_dc^2_\beta + Q_us^2_\beta\right)$,
where the latter term is proportional to the $Z'$ mass squared.


\subsection{Tree-level fine-tuning in the NMSSM+}


Imposing either $\dd{Z}_4^{qq}$ or $\dd{Z}_4^{lq}$
and the approximate $\dd{Z}_3^{HD}$
leads to the low energy NMSSM+ in equation~(\ref{NMSSM+})
which in turn leads to the following 
scalar potential, where we only include fields able to acquire
VEVs, namely $H_{(d,u)3}$ and $S_i$,
\be
{V} &\approx& \lambda^2|S|^2\left(|H_{d}|^2+|H_{u}|^2\right) + |\lambda H_dH_u + k_{333}S_3^2|^2\nn\\
&&\phantom{~} + |k_{222}S_2^2 + 2k_{[221]}S_2 S_1 + k_{[211]}S_1^2|^2\nn\\
&&\phantom{~} + |k_{111}S_1^2 + 2k_{[112]}S_1 S_2 + k_{[122]}S_2^2|^2\nn\\
&&\phantom{~} + \f{g_1^2}{8}\left(|H_{d}|^2-|H_{u}|^2\right)^2 + \f{g_2^2}{8}\left(H_{d}^\dagger\sigma^aH_{d}-H_{u}^\dagger\sigma^aH_{u}\right)^2\nn\\
&&\phantom{~} + \bigg[\lambda A_\lambda S_3H_d.H_u + \f{k_{333}A_{k_{333}}}{3}S_3^3
+ \sum_{\alpha,\beta,\gamma}\f{k_{\alpha\beta\gamma}A_{k_{\alpha\beta\gamma}}}{3}S_\alpha S_\beta S_\gamma + \mbox{c.c.}\bigg]\nn\\
&&\phantom{~} + m_{S_3}^2|S_3|^2 + m_{S_2}^2|S_2|^2 + m_{S_1}^2|S_1|^2 + m_d^2|H_d|^2 + m_u^2|H_u|^2.
\ee


Classically minimising ${V}$ with respect to $s_3^2$ yields
\be
\left\langle\f{\partial{V}}{\partial(s^2)}\right\rangle = 0 &\approx& 4\lambda^2v^2 + 2m_S^2 + 8k\left(\lambda v_dv_u + ks^2\right)\nn\\
&&\phantom{~} + 2\sqrt{2}\lambda A_\lambda\f{v_dv_u}{s} + 2\sqrt{2}kA_ks,
\ee
where $k=k_{333}$ and $s=s_3$, and
classically minimising with respect to $v^2$ yields
\be
\left\langle\f{\partial{V}}{\partial(v^2)}\right\rangle = 0 &\approx&
\bar{g}^2(\cos^2\beta - \sin^2\beta)^2v^2 + 8\lambda \sin\beta \cos\beta\left(\lambda \sin\beta \cos\beta v^2 + ks^2\right)\nn\\
&&\phantom{~} + 4\lambda^2s^2 + 2m_d^2\cos^2\beta + 2m_u^2\sin^2\beta\nn\\
&&\phantom{~} + 4\sqrt{2}\lambda A_\lambda s\cos\beta \sin\beta.
\ee
By design of $\dd{Z}_3^{HD}$ these conditions are approximately independent of $s_{2}$ and $s_1$.
There is no tree-level fine-tuning required for these minimisation conditions to yield
EW scale VEVs since the $Z'$ mass does not appear
and the active singlet VEV $s$ may be taken to be low,
since it is unrelated to the exotic fermion masses, with $\lambda $ reasonably large
in order to yield a large correction to the Higgs boson tree-level mass,
so that the effective $\mu$ term is not too large.
As mentioned in the Introduction, we also expect this model to exhibit less fine-tuning overall
than the NMSSM due the presence of the extra matter which allows 
for larger values of $\lambda$ without violating perturbation theory up to the GUT scale.
The Higgs mass equation~(\ref{eq:hmassNMSSM}) is again relevant, but the presence of
extra matter should allow greater values for $\lambda$ to be perturbative
up to high scales (see e.g. \cite{Masip:1998jc}), increasing the
tree-level Higgs boson mass and ameliorating the need for large loop corrections.

\section{Conclusion}
It is well known that the scale invariant NMSSM has lower fine-tuning than the MSSM, but suffers from
the domain wall problem. In this paper we have 
proposed a new version of the scale invariant NMSSM, called the NMSSM+,
which introduces extra matter
in order to reduce even more the fine-tuning of the NMSSM.
This is not the first time that adding extra matter to the NMSSM to reduce fine-tuning
has been considered, however usually the extra matter that is added is motivated by 
gauge mediated SUSY breaking~\cite{Masip:1998jc}.
In this paper the extra matter descends from an E$_6$ gauge group
and fills out three complete 27-dimensional representations at the TeV scale,
as in the E$_6$SSM. However the $U(1)_N$ gauge group of the E$_6$SSM is broken at a high energy scale 
leading to reduced fine-tuning relative to the fine-tuning in the E$_6$SSM.

One of the motivations for introducing the NMSSM+ is that we have shown that the E$_6$SSM 
as usually realised requires significant tree-level fine-tuning due to experimental
limits on the mass of its $Z'_N$ gauge boson. However, if the 
extra $U(1)_N$ gauge symmetry of the E$_6$SSM is broken
at a high energy scale by extra fields in an approximately $D$-flat
direction, then this relaxes the fine-tuning considerably.
From this point of view, the NMSSM+ may be regarded as the 
E$_6$SSM with $U(1)_N$ gauge symmetry broken at a high energy scale,
with associated scale invariant trilinear singlet couplings.
This then leads to a low energy effective NMSSM+ that resembles the
NMSSM with extra matter. 

The resulting low energy NMSSM+ is summarised in equation~(\ref{WWW}),
which approximates to equation~(\ref{WWWapprox}) (or, equivalently, 
equation~(\ref{NMSSM+})). This low energy NMSSM+ represents a very complete
formulation of the scale invariant NMSSM plus extra matter,
including explicit couplings for the extra matter. Much of the low energy
phenomenology of the extra matter has been discussed in the context of the
E$_6$SSM, however there are some important differences. For one thing, there
is no low energy $U(1)_N$ gauge group or associated $Z'_N$ gauge boson
in the NMSSM+, since this gauge group is broken at a very high energy scale.
Also, all three singlets $S_i$ gain VEVs in the NMSSM+, with $S_3$ being responsible
for the $\mu =\lambda \langle S_3 \rangle $ term, while
$S_{1,2}$ are responsible for $D$ and $\bar{D}$ masses.
This division of labour between the three singlets allows $S_{1,2}$ VEVs to be larger
than the $S_3$ VEV, leading to the $D$ and $\bar{D}$ masses being larger than the 
$\mu = \lambda \langle S_3 \rangle $ term, while keeping $\lambda $ quite large
in order to maximise the tree-level contribution to the Higgs boson mass.

The high energy NMSSM+ in equation~(\ref{NMSSM+3}) provides 
a resolution of the domain wall problem of the NMSSM due to a discrete $R$-symmetry, 
which also stabilises the proton. The renormalisable part of the high energy model in 
equation~(\ref{ren}) contains an explicit sector which breaks the $U(1)_N$ gauge symmetry,
while the non-renormalisable terms lead to trilinear singlet couplings, and other higher order terms
responsible for destabilising the cosmological domain walls.
We gave options for discrete $R$-symmetries that can allow the scenario to be realised
and which forbid rapid proton decay and avoid the domain wall problems.
We also explored approximate flavour symmetries that can suppress flavour changing neutral
currents and naturally allow for a slight difference between the (radiatively induced) scales of
EWSB and of exotic, coloured fermion masses.

Finally we recall there are two equivalent ways of looking at the NMSSM+, namely either as an extension 
of the scale invariant NMSSM with the exotic sector of the E$_6$SSM,
or as an extension of the \essm by the inclusion of cubic
singlet terms (with the $U(1)_N$ broken near the GUT scale).
Either way, the NMSSM+ has some remarkable features compared to the NMSSM or the 
E$_6$SSM. In particular it solves the domain wall problem of the scale invariant
NMSSM via a discrete $R$-symmetry and
is expected to exhibit less overall fine-tuning than either the scale invariant NMSSM
or the E$_6$SSM.

\section*{Acknowledgements}
SFK acknowledges partial support 
from the STFC Consolidated ST/J000396/1 and EU ITN grants UNILHC 237920 and INVISIBLES 289442.

\end{document}